# Tuning the Resonance Properties of 2D Carbon Nanotube Networks towards Mechanical Resonator

Haifei Zhan[1], Guiyong Zhang[2], Baocheng Zhang[3], John M. Bell[1], and Yuantong Gu[1],*

[1]*School of Chemistry, Physics and Mechanical Engineering, Queensland University of Technology, 2 George St, Brisbane QLD 4001, Australia*

[2]*State Key Laboratory of Structural Analysis for Industrial Equipment, School of Naval Architecture, Dalian University of Technology, Dalian 116024, China*

[3]*College of Engineering, Ocean University of China, Qingdao 266071, China*

**\*Corresponding Author:** Professor Yuantong GU

**Mailing Address:** School of Chemistry, Physics and Mechanical Engineering,

        Queensland University of Technology,

        GPO Box 2434, Brisbane, QLD 4001, Australia

**Telephones:** +61-7-31381009         **Fax:** +61-7-31381469

**E-mail:** yuantong.gu@qut.edu.au

**Abstract:** The capabilities of the mechanical resonator-based nanosensors in detecting ultra-small mass or force shifts have driven a continuing exploration of the palette of nanomaterials for such application purpose. Based on large-scale molecular dynamics simulations, we have assessed the applicability of a new class of carbon nanomaterials for the nanoresonator usage, i.e., the single wall carbon nanotube (SWNT) network. It is found that the SWNT networks inherit the excellent mechanical properties from the constituent SWNTs, possessing a high natural frequency. However, although a high quality factor is suggested from the simulation results, it is hard to obtain an unambiguous Q-factor due to the existence of vibration modes in addition to the dominant mode. The nonlinearities resulting from these extra vibration modes are found to exist uniformly under various testing conditions including different initial actuations and temperatures. Further testing shows that these modes can be effectively suppressed through the introduction of axial strain, leading to an extremely high quality factor in the order of $10^9$ estimated from the SWNT network with 2% tensile strain. Additional studies indicate that the carbon rings connecting the SWNTs can also be used to alter the vibrational properties of the resulting network. This study suggests that the SWNT network can be a good candidate for the applications as nanoresonators.

1. Introduction

Recent years have witnessed huge interests from the scientific and engineering communities in fabricating nanoscale mechanical (resonance-based) sensors, whose working principle is to detect the shift in the natural resonance frequency when a foreign mass is attached. The exceptional mechanical properties of the constituent nanomaterials enable the resonance-based sensors to operate at gigahertz or terahertz frequencies, A carbon nanotube (CNT)-based mechanical mass sensor has been reported with a resolution of 1.7 x $10^{-24}$ g, corresponding to the mass of one proton [1]. For the resonator-based mass sensor, the minimum resolvable mass change $\delta m$ is determined by the minimal observable frequency shift $\delta f$ through $\delta m = 2 m_{eff} 10^{-DR/20} \sqrt{\Delta f / \omega Q}$, where $m_{eff}$ is the effective resonator mass, $Q$ as the quality factor, $DR$ and $\Delta f$ as the dynamic range (in units of dB) and measurement bandwidth, respectively, and $\omega$ is the resonance frequency [2]. Achieving a high mass sensitivity requires reducing the effective mass, augmenting the resonant frequency and quality factor, and maximizing the dynamic range.

With these considerations, significant efforts have been devoted to extend the palette of nanomaterials with exceptional mechanical properties and low mass density. Carbon-based nanomaterials, such as CNTs and graphene have received the most intensive attention due to their superior mechanical properties and light weight [3-6]. Recently, another class of 2D carbon-based nanomaterial has been reported [7], which comprises of single wall carbon nanotubes (SWNTs), termed as 2D SWNT networks. Plenty of experimental works have reported the fabrication of randomly aligned 2D SWNT networks, which opens up the avenue to realize the excellent thermal, electrical, mechanical and functionalization properties of CNTs on a macroscopic size scale [8, 9]. A wide range of applications arisen from such networks includes sensors and actuators [10], electronics [11, 12], and energy devices [13, 14].

Regularly aligned 2D SWNT networks can be built by welding X-, Y- or T-like SWNT-based junctions through electron beam under high temperature [15, 16], resulting different hexagonal and orthogonal morphologies. Preliminary studies have examined a wide range of the properties of these SWNT, including the bending rigidity and shear stiffness [17], rupture strain [18], toughness and stiffness [19], equivalent Young's modulus [20, 21], Poisson's ratio [22], and thermal conductivity [23]. It is suggested that these networks inherit the excellent mechanical properties from the constituent CNTs. Together with the porous nature of the structure, i.e., low mass density, and also large surface area, these SWNT networks appear as a promising candidate for the fabrication of nanoresonators.

Therefore, in this work, we assessed the applicability of the SWNT networks for resonator applications through the evaluation of their vibrational characteristics. Note that a previous work has investigated the vibrational properties of the SWNT network under free vibration [24]. However this only discussed the natural frequency and vibration mode of a plate model different from the resonator applications discussed here. In this work, we employed intensive molecular dynamics (MD) simulations to probe the natural frequency of a regularly aligned SWNT network (simplified as a thin beam model) and its quality factor, together with a broad investigation to seek the effective way to tune these resonance properties.

## 2. Computational Methods

To acquire the resonance properties, a series of large-scale MD simulations have been performed on a representative CNT network (Figure 1a), using the open-source *LAMMPS* code [25]. The initial size of the CNT network is chosen as 78×10 nm$^2$ (containing 42432 atoms), and it possesses a hexagonal structure that is comprised of Y-junctions. These Y-junctions are made from (6,6) SWNTs with an angle of 120° between two adjacent neighbour SWNTs (Figure 1c). The widely used reactive empirical bond order (REBO) potential [26] is adopted to describe the C-C interactions. Before applying the actuation to the structure, the conjugate gradient algorithm is applied to obtain the initial equilibrium configuration of the network, and then the sample is equilibrated at 10 K (NPT ensemble, under a pressure of 1 atm) for 500 ps by the Nose-Hoover thermostat [27, 28]. Afterwards, the two ends of the structure are fixed to represent a doubly-clamped working boundary condition and a sinusoidal velocity excitation $v(z) = A\sin(ky)$ is imposed on the structure along the *z*-axis (Figure 1b). Here *A* is the actuation amplitude (equal to 1.0 Å/ps), and *k* is $\pi/L$ with *L* as the effective length of the SWNT network that excludes the two fixed edges. The structure then undergoes free vibration in the NVE ensemble. A time step of 0.5 fs is chosen during the whole simulation. Such resonance simulation approach has been successfully employed to probe the vibrational behaviours of metal nanowires and bi-layer graphene [29, 30].

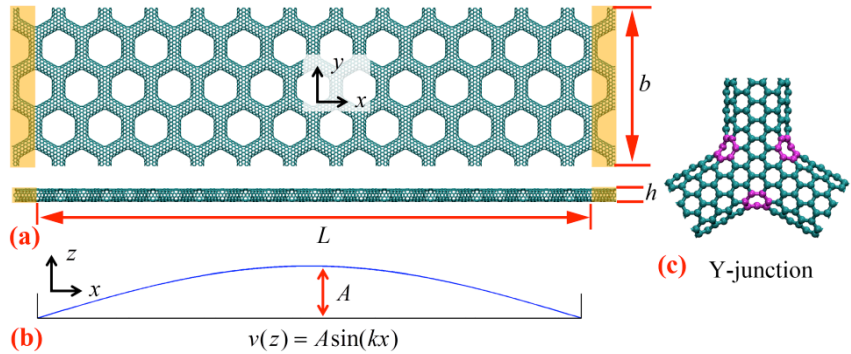

**Figure 1** (a) Schematic view of the hexagonal SWNT network; (b) The profile of the initial actuation; (c) A Y-junction unit for the hexagonal network, denoted as H-A.

Within the energy-preserving NVE ensemble, the energy dissipation during the vibration of the SWNT network will only be resulted from the intrinsic loss, and the loss of potential energy must be converted to the kinetic energy. Thus, the quality factor can be defined as the ratio between the total external energy $E$ and the average external energy loss $\Delta E$ in one radian at the resonance frequency, i.e., $Q = 2\pi E / \Delta E$. Here, the external energy is the difference of the potential energy before and after the application of the initial actuation. Assuming a constant quality factor during vibration, the external energy amplitude $E_n$ (after $n$ vibration cycles) is related to the initio energy amplitude $E_0$ by [31]

$$E_n = E_0(1 - 2\pi / Q)^n \tag{1}$$

Employing Eq. 1 to fit with the recorded trajectory of the external energy, the quality factor can then be derived. We should note that the absolute $Q$ value can vary with the applied calculation approach [32], which however will not change the scaling behaviour, which is the focus of this paper. Based on the time history of the external energy, we can also directly extract the vibrational frequency $f_n$ by fast Fourier transformation [33]. Further, the studied SWNT networks can be approximated as a continuous thin beam model, i.e., the vibrational frequency can be related to Young's modulus $E$ through the classical Euler-Bernoulli beam theory as [34]

$$f_n = \frac{\omega_n}{2\pi L^2} \sqrt{\frac{EI}{\rho S}} \tag{2}$$

Here $L$ and $\rho$ are the length and mass density of the beam, respectively. $S$ is the cross-sectional area. $\omega_n$ is the eigenvalue obtained from the characteristic equations.

### 3. Results and Discussions

*3.1 Vibrational behaviours*

Figure 2a presents the time history of the external energy for the vibration system, from which the trajectory of the amplitude exhibits a waved shape with the increasing simulation time. This phenomenon signifies the existences of multiple vibrational modes, i.e., a nonlinear vibration characteristic. As evidenced from the corresponding frequency spectrum in Figure 2b, there exist several resolvable frequency peaks beside the major peak. These extra frequency components arise from several origins, including the random temperature variation during simulation (a white noise), higher vibrational modes (the nonlinearity as seen in lower of Figure 2c), lateral vibration modes and also the so-called edge vibration modes due to the free edges of the sample (which is also observed from the resonance of a pristine graphene [35]). Despite of the waved shape, the external energy amplitude does not show a clear decreasing trend, indicating the examined structure has a high quality factor, though the presence of those extra vibrational modes prevents an accurate estimation of its value. We should note that different filters (such as lower and high-pass filters) could be used to isolate the dominant vibration mode, however, the resulting $Q$ value will highly rely on the filters being used. Therefore, we only estimate the $Q$ factor for the cases that do not rely much on the smoothing processing in the following.

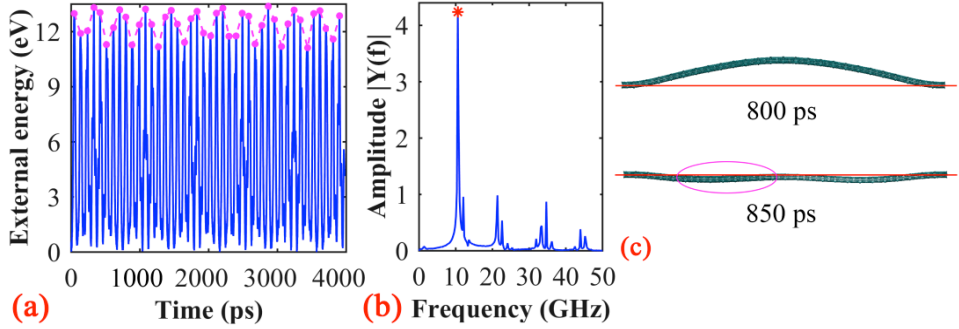

**Figure 2** Simulation results from the SWNT network: (a) the time history of the external energy; (b) the corresponding frequency spectrum; (c) the atomic configurations during vibration.

On the other hand, by tracking the atomic configurations, the SWNT network is found to exhibit a sinusoidal profile during vibration (upper of Figure 2c), which corresponds to the first-order vibration mode of a doubly clamped continuum beam model. That is, the dominant frequency peak, estimated as 10.6 GHz from Figure 2b, is actually related to the first-order natural frequency of the structure. According to the double-angle formulas, the structure has a natural frequency of 5.3 GHz, half of that obtained from the external energy. With this value, the effective Young's modulus of the SWNT network can be calculated by approximating the moment of inertia as $I = bh^3/12$ (with $b$ and $h$ as the width and height of the network, respectively). Considering the effective mass density of the model as $\rho = Nm_c/V$ (here $N$ is the total atom number, $m_c$ is the carbon atom mass which equals $1.67\times10^{-27}$ kg and $V$ is the system volume), the effective Young's modulus is then calculated as 134.5 GPa following Eq. 2, which is comparable with the results obtained from the tensile deformation [22].

### *3.2 Initial actuation and temperature*

Above results suggest that the SWNT network could be a good platform for the nanoresonator, providing that the interferential vibration modes can be well suppressed. One critical factor that affects the vibrational behaviours of the SWNT network is the initial actuation, which can be observed from our simulations with the actuation amplitudes of 0.4, 0.6, 0.8, 1.0, 1.2 and 1.4 Å/ps. As illustrated in Figure 3a, weaker actuation means smaller external energy amplitude, which can greatly avoid the excitation of higher vibration modes, whereas the external energy suffers more obvious disturbance from the temperature variation (see inset of Figure 3a). In the opposite, stronger actuation can suppress the temperature impacts, but trigger higher vibration modes (see Figure 3d and its inset). Along with the direct impacts on the external energy profile, the actuation amplitude also exerts remarkable impacts on the effective natural frequency of the vibration system. A continuous increase of the dominant frequency from 3.6 to 6.0 GHz is observed while the actuation amplitude increases from 0.4 to 1.4 Å/ps (see Figure 3e). Such observation is in line with the theoretical analysis [36] that the natural frequency increases with the increase of the deflection, in the order of $(w_{max}/h)^2$ (here $w_{max}$ is the maximum deflection of the structure).

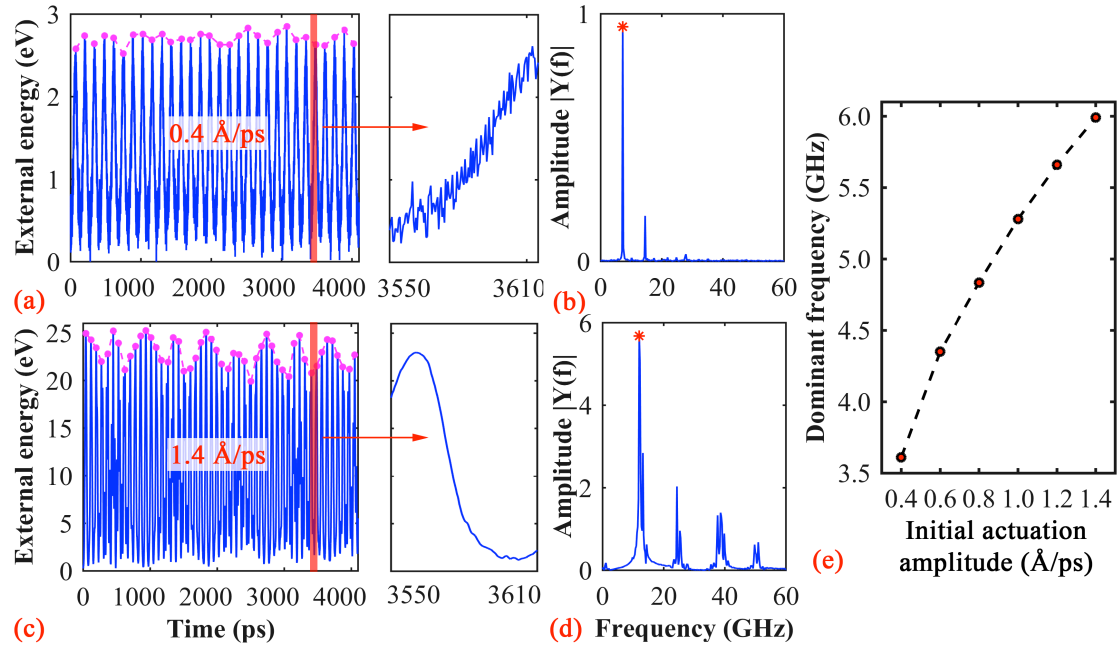

**Figure 3** Simulation results from the structure under different actuation amplitude *A*: (a) the time history of the external energy with *A* as 0.4 Å/ps ; (b) the corresponding frequency spectrum; (c) the time history of the external energy with *A* as 1.4 Å/ps; (d) the corresponding frequency spectrum; insets in (a) and (c) shows a zoom in profile of the external energy during 3550 and 3610 ps; and (e) the dominant resonant frequency as a function of the initial actuation amplitude.

Another factor that greatly influences the vibrational behaviours of the SWNT network is the temperature. With the same initial actuation of 1.0 Å/ps, the vibration of the structure under the temperature of 10, 50, 100, 150, 200, 250 and 300 K has been reproduced. Figure 4 illustrates the simulation results under the temperature of 100 and 300 K, which shows a largely different external energy profile for the vibration system, although the same actuation has been applied. Recall the results in Figure 2a (at 10 K), higher temperature will lead to larger external energy amplitude, which can be explained as higher system temperature means larger random temperature variation during simulation. For instance, the system temperature varies from 10.0 ~ 12.55 K under the 10 K initial temperature after the introduction of the actuation, while it varies from 297.8 ~ 305.5 K under 300 K, corresponding to 2.5 and 7.7 K temperature variations, respectively. Excluding the same amount of energy added by the initial actuation, this increased temperature variation behaves as an enhanced white noise to the recorded external energy, and thus makes it harder to derive a valid quality factor for the system. Also, it is seen from the corresponding frequency spectrum that the relative amplitude of the dominant frequency peak is smaller at higher temperature, indicating larger impacts from other vibrational modes.

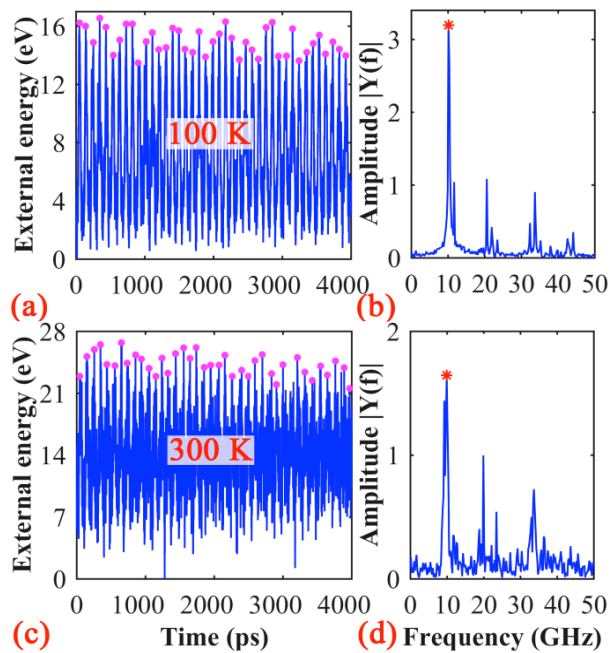

**Figure 4** Simulation results from the structure under different temperature: (a) the time history of the external energy at 100 K; (b) the corresponding frequency

spectrum; (c) the time history of the external energy at 300 K, and (d) the corresponding frequency spectrum.

### *3.3 Axial strain enhancement*

Previous discussions have shown that the initial actuation and temperature have significant impacts on the vibrational behaviours of the SWNT network and lead to nonlinear vibration characteristics, which prevents a solid isolation of the quality factor. Continue efforts are then focused on how to suppress these nonlinearities. For this purpose, we apply a constant strain rate to stretch the network to the strain of 0.02 prior imposing the actuation (1.0 Å/ps). Figure 5a shows the testing results from the structure with a 0.02 pre-tension. Surprisingly, the amplitude of the external energy has preserved a clear linear and gradual decreasing trend, which has led to an extremely high *Q*-factor, ~$1.08 \times 10^9$ (fitted with Eq. 1). In the meanwhile, only the dominant frequency has been resolved as seen in the corresponding frequency spectrum, indicating that the strain can effectively suppress the extra vibrational modes that deteriorate the resonance properties of the SWNT network.

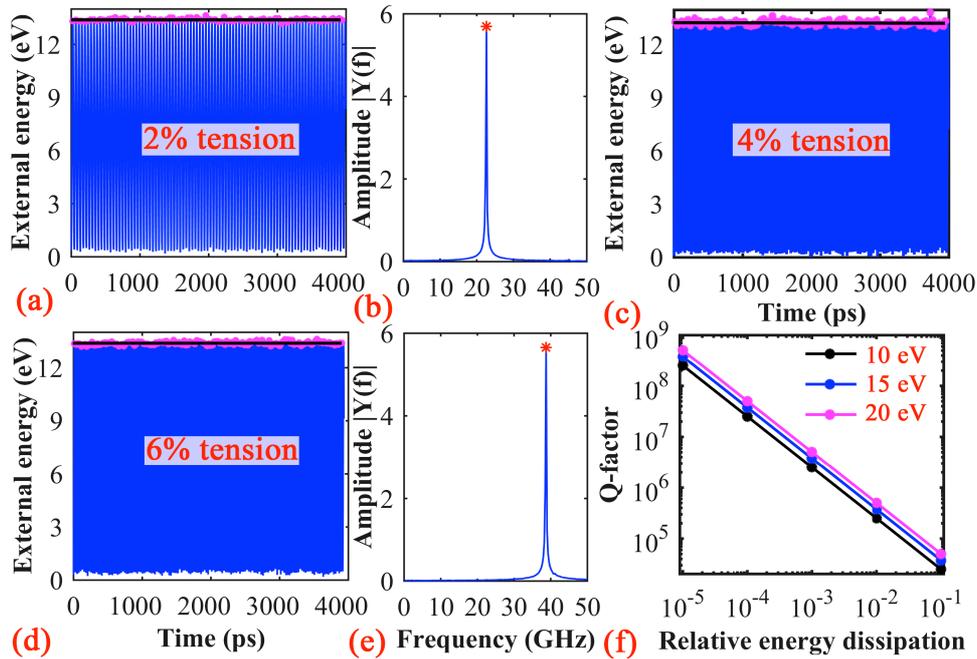

**Figure 5** Simulation results from the structure with pre-tensile strain: (a) the time history of the external energy at the strain of 2%; (b) the corresponding frequency spectrum; (c) the time history of the external energy at the strain of 4%; (d) the time history of the external energy at the strain of 6%; (e) the corresponding frequency spectrum; and (f) the sensitivity of the *Q*-factor on the relative energy dissipation while the initial external energy changes from 10 to 20 eV.

Similar results have also been obtained by examining the structure under larger tensile strain, i.e., 0.04, 0.06 and 0.08 (the SWNT network is still under elastic deformation at these strains). However, it is observed that comparing with the results from the structure with 2% strain, the trajectory of the external energy experiences much more apparent fluctuations (see Figure 5c and 5d), which again induces a wide variation to the estimated $Q$-factor. Using Eq. 1, the $Q$-factor for the structure with 4% and 6% strain is approximated as $3.85 \times 10^5$ and $1.10 \times 10^9$, respectively (the $Q$ for the structure with 8% strain is not calculated as the external energy shows relatively large fluctuations). Surprisingly, the $Q$-factor does not show an explicit relation with the tensile strain. It is worthy to emphasis that the $Q$ is derived by fitting with the external energy amplitude, i.e., the amplitude fluctuations would significantly influence its absolute value. Figure 5f shows the sensitivity of the $Q$-factor under different initial external energy (from 10 to 20 eV) while the relative energy dissipation changes from $10^{-1}$ to $10^{-5}$ (by assuming 40 vibration circles according to the obtained simulation results). Here, the relative energy dissipation is defined as $(E_0 - E_n)/E_0$. Generally, the $Q$ will increase significantly with the decrease of the energy dissipation, and a $10^{-1}$ relative energy dissipation would indicate a $Q$ in the order of $10^4$ to $10^5$. Figure 5f indicates that the SWNT network with different initial strains possesses a relatively high $Q$-factor regardless of the numerical error comes from the fitting process. It is thus yield to the conclusion that the tensile strain can help the 2D SWNT network to achieve a high $Q$-factor and also effectively suppress the nonlinearity, in consistent with previous works on graphene [3, 37].

Besides the enhancement to the $Q$-factor, the continuing increase of the first-order natural frequency has also been observed, which is in line with the theoretical models. According to the continuum mechanics, the first-order natural frequency of a beam under free vibration is related to the initial/residual tensile force $F_0$ by [38]:

$$f_t = \frac{2\pi}{L^2} \sqrt{\frac{EI}{3\rho A}\left(1 + \frac{L^2 F_0}{4\pi^2 EI}\right)} \quad (3)$$

In other words, the relative first-order natural frequency $f_{re} = f_t / f_0$ ($f_0$ as the frequency while $F_0 = 0$) is related to the initial axial strain $\varepsilon_0$ by $f_{re} = \sqrt{1 + C\varepsilon_0}$, where $C$ is a constant for a given beam. As illustrated in Figure 6, the estimated natural frequency under the examined pre-tension values follows well with such relationship.

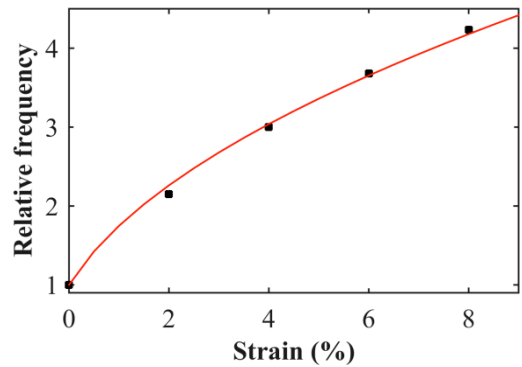

**Figure 6** The relative first-order natural frequency as a function of pre-tensile strain, red line is the fitted curve.

*3.4 Structural Impacts*

Before concluding, we also investigated how the resonance properties of the SWNT network can be altered through the change of its geometry structure. Basing on previous *ab initio* study [39], there are a variety of stable 2D SWNT networks can be constructed. Figure 6a shows another three junction units that we choose to build the network. Under the same simulation settings (i.e., 10 K, actuation amplitude of 1.0 Å/ps), the vibrational properties are observed to vary with different structures.

Generally, it is found that the two hexagonal SWNT networks possess a similar waved shape of external energy amplitude (see Figure 2a and Figure 7b). Whereas, for the two orthogonal structures, a largely different external energy amplitude has been observed. As illustrated in Figures 7d and 7g, the amplitude of the external energy for one of the orthogonal structures (O-A) exhibits a clear and quick decreasing trend, which leads to a *Q*-factor around 390 (according to Eq. 1). Such observation indicates that the connection carbon rings also play a critical role on the vibrational properties of the SWNT networks. We should note that all considered SWNT networks are anisotropic along the two lateral directions, that is, the mechanical properties are different in different directions. Such anisotropic properties will induce different vibrational properties while the model has different orientations. The example results are given in Figures 7e and 7h for the two orthogonal structures while the connecting carbon rings are aligned along the *y*-direction (rotate 90°). As is seen, in this circumstance, the amplitude of the external energy of the O-A structure decreases slowly, indicating a much better *Q*-factor (estimated as 10790). In the other hand, despite the nonlinearity, the amplitude of the external energy for the O-B structure exhibits a clear decreasing tendency.

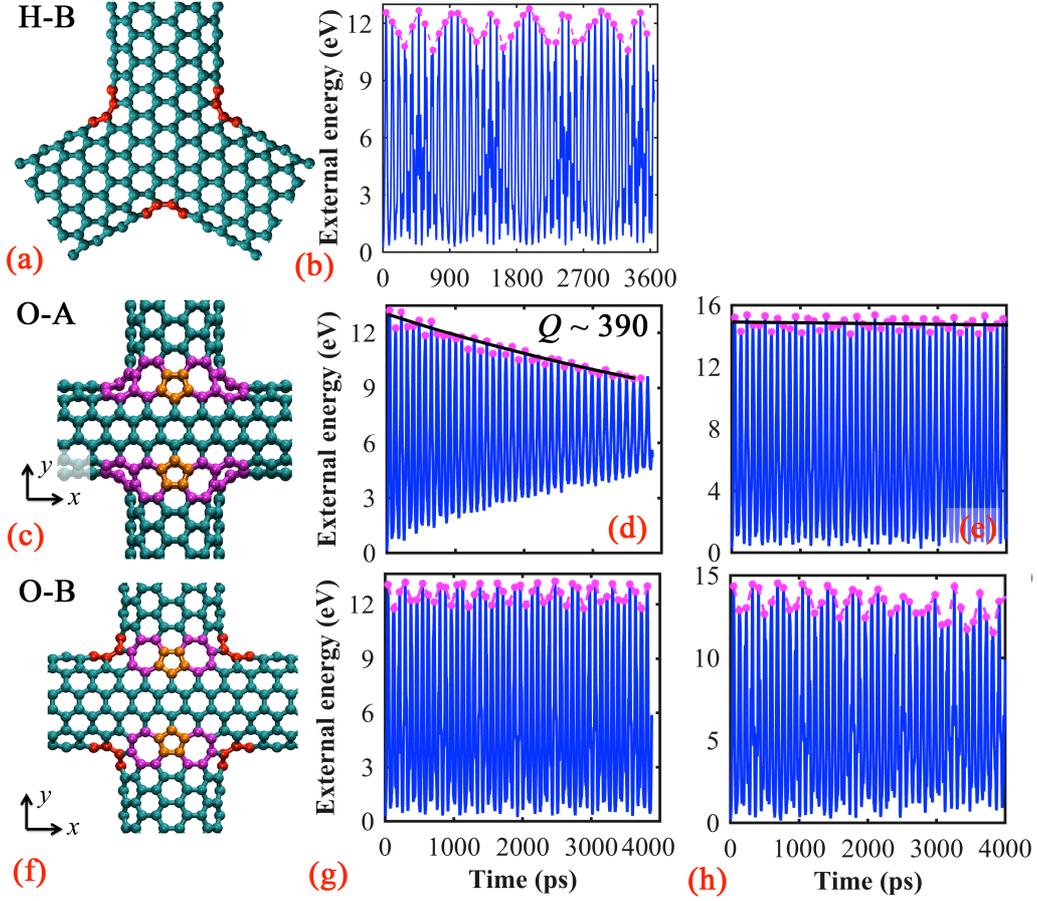

**Figure 7** The simulation results for different SWNT networks. (a) The H-B unit junction, and (b) the corresponding time history of the external energy; (c) the O-A unit junction, and the corresponding time history of the external energy while the connecting carbon rings are aligned along: (d) the *x*-axis and (e) the *y*-axis, black line represents the fitting from Eq. 1; (f) the O-B unit junction, and the corresponding time history of the external energy while the connecting carbon rings are aligned along: (g) the *x*-axis and (h) the *y*-axis. Here *x*-axis is the length direction of the structure.

In addition, due to the different connection carbon rings being used to weld the SWNTs, the networks have different effective mass density and mechanical properties. Interestingly, we found that these examined structures have a close first-order natural frequency around 5.5 GHz, which leads to a similar effective Young's modulus in the range of 130 ~ 213 GPa as listed in Table 1. These results are well in agreement with the results obtained from the tensile testing [22].

**Table 1** The properties of different 2D SWNT networks. Not that the O-A and O-B networks are the structure with connecting carbon rings aligned along *x*-axis.

| Properties\Model type | H-A | H-B | O-A | O-B |
|---|---|---|---|---|
| First-order natural frequency (GHz) | 5.3 | 5.8 | 5.5 | 5.9 |
| Effective mass density (kg/m$^3$) | 101.3 | 113.1 | 115.6 | 123.0 |
| Effective Young's modulus (GPa) | 134.5 | 212.8 | 154.9 | 213.6 |

## 4. Conclusions

Based on a series of MD simulations, we investigated the applicability of the 2D carbon network for the usage of a nanoresonator through a broad discussion on the resonance properties of a representative SWNT network. It is found that the SWNT network possesses a high natural frequency due to the excellent mechanical properties inherited from the constituent SWNTs. However, although a high quality factor is suggested from the simulation results, it is hard to obtain a solid estimation due to the existences of other vibration modes. Further testings show that the temperature and initial actuation exert significant impacts on the vibrational behaviours the SWNT networks, and these extra vibration modes always accompany. It is found that these extra vibrational modes can be effectively suppressed through the introduction of axial strain. An extremely high quality factor in the order of $10^9$ is estimated from the SWNT network with 2% tensile strain. Additional studies reveal that the connection carbon rings that weld the SWNTs can also be used to alter the vibrational properties of the resulting network. This study establishes a first understanding of the vibrational properties of the SWNTs network, which should shed lights on its future applications in the nanoresonator-based devices.

**Acknowledgement**

Zhan is grateful to Dr Zhou from Hefei University of Technology (China) for providing the advices to construct the models. Supports from the ARC Discovery Project (DP130102120), the Recruitment Program of Global Young Experts (China) and Fundamental Research Funds for the Central Universities (DUT14RC(3)002, China), the National Nature Science of Foundation of China (Grant No. 51275487), the High Performance Computer resources provided by the Queensland University of Technology are gratefully acknowledged.